\documentclass[preprint,showpacs,preprintnumbers,amsmath,amssymb]{revtex4}

\usepackage{graphicx}
\usepackage{dcolumn}
\usepackage{bm}

\begin{document}

\title {Spin Transfer from the point of view of the ferromagnetic 
degrees of freedom}
\author{J.-E. Wegrowe} 
\affiliation{Ecole Polytechnique, LSI, CNRS and CEA/DSM/IRAMIS, Palaiseau 
F-91128.} 


\begin{abstract}
Spintronics is the generic term  that describes magnetic systems
coupled to an electric generator, taking into account the spin
attached to the charge carriers.  For this topical review of {\it Spin
Caloritronics}, we focus our attention on the study of {\it
irreversible processes} occuring in spintronic devices, that involve
both the spins of the conduction electrons and the ferromagnetic
degrees of freedom.  The aim of this report is to clarify the nature
of the different kinds of power dissipated in metallic ferromagnets
contacted to an electric generator, and to exploit it in the framework
of the theory of mesoscopic non-equilibrium thermodynamics.  The
expression of the internal power (i.e. the internal entropy production
multiplied by the temperature) dissipated by a generic system
connected to different reservoirs, allows the corresponding kinetic
equations to be derived with the introduction of the relevant
phenomenological kinetic coefficients.  After derivation of the
kinetic equations for the ferromagnetic degrees of freedom (i.e. the
Landau-Lifshitz equation) and the derivation of the kinetic equations
for the spin-accumulation effects (within a two channel model), the
kinetic equations describing spin-transfer are obtained.  Both
spin-dependent relaxation (usual spin-accumulation) and
spin-precession in quasi-ballistic regime (transverse
spin-accumulation) are taken into account.  The generalization of the
Landau-Lifshitz equation to spin-accumulation is then performed with
the introduction of two potential energy terms, that are
experimentally accessible.

\end{abstract}

\pacs{75.40.Gb,72.25.Hg,75.47.De \hfill}

\maketitle

\section{Summary}

The approach of spintronics adopted here is that of Non-Equilibrium
Thermodynamics \cite{DeGroot,Prigogine,Stueckelberg,Smith,Parrott}
applied at the Mesoscopic scale \cite{Vilar,Regera} (MNET).  The
analysis is based on the expression of the entropy production, i.e. on
the expression of the power dissipated through the different
relaxation mechanisms that characterize the system.  The theory is
adapted to the description of a plurality of out-of-equilibrium
sub-systems exchanging energy, in which the role of environmental
microscopic degrees of freedom are reduced to transport coefficients
(damping, diffusion coefficients, conductivity, thermoelectric power,
gyromagnetic ratio, and other Onsager coefficients) for the collective
variables under consideration.  The description holds at the
mesoscopic scales, under the hypothesis of local equilibrium extended
to internal degrees of freedom.  The aim of this report is to propose an
application of MNET to metallic spintronic devices that includes
uniform ferromagnetic degrees of freedom explicitly.

Before presenting the detailed derivation of the kinetic equations in
the forthcoming sections II, III and IV, let us first summerize the
general scheme of the report.  In this introductory section, the usual
thermoelectric effect is first presented, and extended to the case of
a bi-valuated internal variables: this is the two-channel model.  An
analogous approach is then performed in the space of the ferromagnetic
degrees of freedom $\Sigma$. 
The coupling between the $\Sigma$ space and the internal degree of
freedom of the electronic sub-system leads us to the spin-transfer
kinetic equations.  The detailed treatment of the ferromagnetic
transport is given in section II, the detailed treatment of the two
channel model of electric transport is performed in section III, and
the coupling between both subsystems, i.e. the spin-transfer 
\cite{Berger,Slon}, is presented
in the last section.

\subsection{Thermoelectric effects}

The internal power dissipated by a system is given by the internal
entropy production $\frac{dS^{i}}{dt}$ (where $S$ is the entropy of
the system) multiplied by the temperature $T$.  In the case of
electric charges moving in one dimension $z$ in a wire of section
unity (i.e. the wire is contacted to two reservoirs of
electric charge) and maintained at uniform temperature , the power dissipated
inside the system is given by the Joule heating, i.e. the product of
the electric current by the electric field $\mathcal{E}$:

\begin{equation}
T \frac{dS_{i}^{e}}{dt} = J^{e}.\mathcal{E} =
- J^{e} . \frac{1}{e} \frac{\partial  \mu^{e}}{\partial z}
\end{equation}

where $J^{e}$ is the electric current, $e$
the absolut value of the electric charge, and $\mu^{e}$ is the electrochemical
potential. The electric field  is given by $\mathcal{E}
= -\frac{1}{e} \frac{\partial \mu^{e}}{\partial z} $.

The application of the second law of thermodynamics $\frac{dS_{i}}{dt} 
\ge 0$ leads us to define a 
first positive Onsager coefficient $\sigma$ (which is a function of the 
state variables) in order to built a
positive quadratic form. The electric current writes:

\begin{equation}
J^{e} =  -\frac{\sigma}{e} \frac{\partial \mu^{e}}{\partial z} 
\end{equation}

which is Ohm's law and  $\sigma$ is the electric conductivity.

On the other hand, the power 
dissipated inside a wire contacted to two heat reservoirs is the 
product of the heat flow $J^{Q}$ and the 
conjugated force $\frac{\partial}{\partial z} \left (\frac{1}{T} \right )$ 
multiplied by $T$:

\begin{equation}
T \frac{dS_{i}^{Q}}{dt} = 
T J^{Q} . \frac{\partial}{\partial z} \left ( \frac{1}{T} \right)
\end{equation}
 
The Fourier's law is deduced from the second law of thermodynamics after 
introducing a positive Onsager coefficient $\kappa$:

\begin{equation}
J^{Q} =  -\kappa \frac{\partial T}{\partial z}
\end{equation}

where $\kappa$ is the thermal conductivity.

If both electric and heat reservoirs are contacted to the same wire, the 
internal power dissipated is:

\begin{equation}
T \frac{dS_{i}}{dt} = 
-J^{e} . \frac{1}{e}\frac{\partial \mu^{e}}{\partial z} + 
T J^{Q}. \frac{\partial}{\partial z} \left (\frac{1}{T} \right )
\end{equation}
  
Now, beyond Ohm's law and Fourier's law, the two currents are 
coupled through the relevant Onsager thermoelectric 
{\it cross-coefficients},

\begin{equation}
\left\{
\begin{aligned}
J^{e} &= -\frac{\sigma}{e} \frac{\partial
\mu^{e}}{\partial z} + \mathcal{S} \sigma \frac{\partial
T}{\partial z}\\
J^{Q} &=  \frac{\mathcal{S} T \sigma}{e} \frac{\partial \mu^{e}}{\partial z}  
- \left ( \kappa + T \mathcal{S}^{2} \sigma \right) \frac{\partial T}{\partial z}
\end{aligned}
\right.
\label{TEP0}
\end{equation}

where the Onsager cross-coefficients are expressed with the help if 
the Seebeck
coefficient $\mathcal{S}$, defined at zero electric current by the 
relation $\mathcal E = \mathcal{S} \, \partial T / \partial z$ and
the conductivities $\sigma$ and $\kappa$. Note that according to the Onsager
reciprocity relations, the Peltier coefficient $ \Pi$, defined without 
electric field by the relation $J_{Q} = \Pi J^{e}$, verify to the 
relation $\Pi = T\mathcal{S}$. 
The existence of the cross-coefficients is justified by the fact that
the charge carriers are also contributing to the transport of heat, or
inversely, the transport of heat is contributing to the transport of 
electric charges. If a detailed microscopic theory is possible in the case of 
the Ohm's law and the Fourier's law, this is no longer the case in 
general for thermoelectric effects or other cross-effects. However, 
the knowlege of the detailed mechanisms of heat transport due to 
electric carriers is not necessary in order to derive Eq. (\ref{TEP0}). This 
justifies the interest of a thermokinetic phenomenological appoache 
applied to spin caloritronics, for which the underlaying relaxation 
mechanisms are 
also not well known \cite{Hickey,Damping}.

\subsection{Two-channel relaxation and spin-accumulation}

Let us assume that the ensemble of electric charges is composed of two
different populations.  The difference is introduced through an
internal degree of freedom, restricted here to a bi-valuated variable
that takes the value $\alpha$ and $\gamma$.  In the context of
semiconductor physics, the two channel model was introduced in order
to describes the transport of both electrons and holes
\cite{Smith,Parrott}.  In the context of the usual spin-accumulation 
effect
due to spin-flip relaxation, the two channels account for the
spin up or spin down attached to the conduction electrons: $\alpha =
\uparrow$ and $\gamma =\downarrow$ \cite{Johnson,Wyder,ValetFert,PRB00}. 
However, from the point of view adopted here, this scheme should be
generalized to a band structure in order to account for the $s-d$
relaxation mechanisms, that are responsible for the coupling between
the spin of the conduction electron (mainly $s$ electron band) and the
ferromagnetic degrees of freedom (related to $d$ electron band)
\cite{Revue}.  In this case, electronic transport is also
spin-dependent because the $d$ band is full for majority spins 
(e.g. $\uparrow$) in usual 3d metallic
ferromagnets \cite{Mott}.

However, without entering into the complexity of the spin-dependent relaxation
mechanisms, it is easy to generalize Ohm's law with adding the
parameters $\alpha$ and $\gamma$ to the transport coefficients.  A third 
kinetic equation should be
introduced in order to take into account the power dissipated by the
$\alpha \rightarrow \gamma $ relaxation (spin-flip relaxation or $s-d$
relaxation).  This relaxation is formally equivalent to a chemical
reaction, driven by the chemical affinity $\Delta \mu = \mu_{\alpha} -
\mu_{\gamma}$ \cite{PRB00}.  Indeed, the power dissipated by the system 
reads
then:

\begin{equation}
T \frac{dS_{i}^{e}}{dt} =
- J_{\alpha}^{e} . \frac{1}{e} \frac{\partial  \mu_{\alpha}^{e}}{\partial 
z} -  J_{\gamma}^{e} . \frac{1}{e} \frac{\partial  \mu_{\gamma}^{e}}{\partial 
z} + \dot \Psi \Delta \mu
\label{PowerGMR}
\end{equation}

where the flux of particles relaxing from one channel to the other
(relaxation occurring in the space of the internal degrees of freedom)
is given by $\dot \Psi$.  The corresponding kinetic equations are
deduced, after introducing a supplementary Onsager coefficient $L$.

\begin{equation}
\left\{
\begin{aligned}
J^{e}_{\alpha } &= -\frac{\sigma_{\alpha}}{e} \frac{\partial
\mu^{e}_{\alpha}}{\partial z} \\
J^{e}_{\gamma } &= -\frac{\sigma_{\gamma}}{e} \frac{\partial
\mu^{e}_{\gamma}}{\partial z} \\
\dot \Psi &= L \Delta \mu
\end{aligned}
\right. \label{OnsagerGMR}
\end{equation}

The set of equations Eqs (\ref{OnsagerGMR}) is sufficient and
necessary in order to describe, in the stationary regime, spin-accumulation 
effects or any
non-equilibrium contribution to the resistance due to $\Delta \mu$ occurring at an
interface. The corresponding effects (spin
accumulation, giant magnetoresistance, etc) will be discussed in 
Section III. Onsager cross-coefficients may also be added at this stage of
the analysis \cite{Ansermet}.  

It is important to note the introduction of the parameter $\Delta \mu$ in the irreversible processes 
described in Eq. (\ref{PowerGMR}): in spintronics,  $\Delta \mu$ is called "spin accumulation" 
and it was first introduced in this context by Van Kempen et al. 
\cite{Wyder}. This parameter plays the role of the pumping 
force that is responsible for the out-of-equilibrium relaxation 
occurring from 
one channel to the other. It will be also responsible for 
spin-transfer as described at the end of this report.

Accordingly, it is convenient
to rewrite Eq. (\ref{OnsagerGMR}) as a function of the variable $\Delta \mu$. 
Let us 
define  the conductivity asymmetry by the parameter $\beta$ such
that $\beta = \frac{\sigma_{\alpha} - \sigma_{\gamma}}{\sigma_{0}}$ and
the mean conductivity $ 2 \sigma_{0}= \sigma_{\alpha} + \sigma_{\gamma}$.  On the other hand, the
spin-polarized electric current is $\delta J^{e} = J^{e}_{\alpha} -
J^{e}_{\gamma} $ and the spin-independent current is $J^{e}_{0} = 
J^{e}_{\alpha} + J^{e}_{\gamma} $.  In this new system of equations, the Onsager matrix
re-writes:

\begin{equation}
\left( \begin{array}{c}
\delta J^e \\
J^{e}_{0} \\
\dot \Psi 
\end{array} \right)
=  \left( \begin{array}{ccc}
                        \sigma_0 & \beta \sigma_0 & 0\\
                       \beta \sigma_0 & \sigma_0 & 0 \\
			 0 & 0 & L
                        \end{array} \right)
                        \left( \begin{array}{c}
\frac{-1}{e} \frac {\partial{\Delta 
{\mu}^{e}}}{\partial {z}}\\
\frac{-1}{e} \frac{\partial \mu^{e}_{0}}{\partial z}\\
\Delta \mu \\
 \end{array} \right)
\label{OnsagerElec00}
\end{equation}

where $\mu_{0} = \mu_{\alpha} + \mu_{\gamma}$.

The generalization of the
thermoelectric effect to the two channel case is straightforward,
assuming that the electrons are thermalized (i.e. the temperature is
the same for each channel):

\begin{equation}
\left\{
\begin{aligned}
J^{e}_{\alpha } &= -\frac{\sigma_{\alpha}}{e} \frac{\partial
\mu^{e}_{\alpha}}{\partial z} + \mathcal{S_{\alpha}} \sigma_{\alpha} \frac{\partial
T}{\partial z} \\
J^{Q}_{\alpha} &= - \mathcal{S}_{\alpha} T \sigma_{\alpha} \frac{\partial 
\mu_{\alpha}^{e}}{\partial z}
- \left ( \kappa_{\alpha} + T \mathcal{S}^{2}_{\alpha} \sigma_{\alpha} \right) \frac{\partial T}{\partial z}\\
J^{e}_{\gamma } &= -\frac{\sigma_{\gamma}}{e} \frac{\partial
\mu^{e}_{\gamma}}{\partial z} + \mathcal{S_{\gamma}} \sigma_{\gamma} \frac{\partial
T}{\partial z} \\
J^{Q}_{\gamma} &= - \mathcal{S}_{\gamma} T \sigma_{\gamma} 
\frac{\partial \mu^{e}_{\gamma}}{\partial z} 
- \left ( \kappa_{\gamma} + T \mathcal{S}^{2}_{\gamma} \sigma_{\gamma} \right) \frac{\partial T}{\partial z}
\end{aligned}
\right. \label{OnsagerMTEP}
\end{equation}

where $S_{i}$ and $\kappa_{i}$, $i=\{\alpha,\gamma \}$, are
respectively the Seebeck and Fourier coefficient of each channel. 
However, in this report, we will not further investigate this set of
equations.  Some consequences in relation with experiments have been
investigated and reported \cite{Sakurai,
Piraux,Shi,Shi2,Fukushima,Gravier,Santi,MTEP,SpSeebeck,WegIEEE}.  The set of equations
(\ref{OnsagerMTEP}), or other theoretical descriptions beyond the two channel
model \cite{BauerTEP,Saslow,Kovalev} investigated the caloritronic properties 
of spintronic systems.  In
contrast, the goal of this report is to describe the transport 
properties of the ferromagnetic
degrees of freedom and the consequences of its interaction with spin-dependent electric
sub-systems.

\subsection{Introduction of the ferromagnetic degrees of freedom}

Let us introduce the ferromagnetic degrees of freedom.  This observable 
is defined in a physical space that is not the usual space $\Re$ considered 
above, but a space of magnetic moments $\Sigma$. This space should be 
considered as a space of internal degrees of freedom, in the same sense 
as the bi-valuated variable \{$\alpha$,$\gamma$ \} of the two channel 
model (and in the same sens as the spin space in quantum mechanics).
The power dissipated by the magnetic system is then given by the flux $\vec J_{0}^{F}$
of magnetic moments (defined in the corresponding vectorial space 
$\Sigma$) multiplied by the magnetic force:

\begin{equation}
    T \frac{dS_{F}^{i}}{dt} = \vec J_{0}^{F} \,. \vec \nabla_{\Sigma} 
    \mu^{F}_{0}
   \label{Sferro}
\end{equation}

where $\mu^{F}_{0}$ is the ferromagnetic chemical potential and $\vec 
\nabla_{\Sigma}$ is the gradient defined in the space $\Sigma$ (see 
Section III). The application of the second law of thermodynamics leads 
us to 
introduce the positive Onsager matrix $\bar{\mathcal L}_{0}$ such that

\begin{equation}
    \vec J_{0}^{F} = \bar{\mathcal L}_{0} \, \vec \nabla_{\Sigma} 
    \mu^{F}_{0}
   \label{Jferro0}
\end{equation}

This kinetic equation for the ferromagnetic degrees of freedom is 
actually the simplest formulation of the Landau-Lifshitz 
equation for the magnetization $\vec M$ (see 
Section II below). 

By analogy with the thermoelectric effect, it is tempting to formally introduce 
a gradient of temperature {\it in the corresponding 
configuration space}. We expect then the existence of a 
supplementary force acting on the magnetization:

\begin{equation}
\left( \begin{array}{c}
 \vec J^{F} \\
 \vec J_{Q}^{F} \\
\end{array} \right)  = 
 \left( \begin{array}{cc}
 \bar{\mathcal L} &  \bar S^{F} \\
 \bar \Pi^{F} &  \bar \lambda \\
\end{array} \right)
 \left( \begin{array}{c}
 \vec \nabla_{\Sigma} \mu^{F} \\
 \vec \nabla_{\Sigma} T \\
\end{array} \right)
\label{TemSigma}
\end{equation}

where $\bar S^{F}$, $\bar \lambda$  and  $\bar \Pi^{F} $ are 
arbitrary Onsager matrices that formally generalises the Seebeck coefficient, the 
thermal conductivity, and the Peltier coefficient in the $\Sigma$ space.
The question is to understand the physical meaning of a temperature
 gradient in the configuration space of the magnetization.  This
 situation is analogous but very different from the case of two thermostats of
 different temperatures localized in two places (in the $\Re$
 space).  The question about the physical signification
 of a quantity like $\vec \nabla_{\Sigma} T $ is not trivial. 
 
 However, the situation described in Eq. (\ref{TemSigma}) is rather 
 similar to the result obtained at the end of the report, providing 
 that the 
 effective temperature gradient $\vec \nabla_{\Sigma} T$ is replaced by 
 the voltage drop due to spin-accumulation:  $\Delta \mu \approx \int 
 \frac{\partial \Delta \mu }{dz}dz $. A supplementary force is acting 
 on the ferromagnet.
 
 Indeed, let us consider now the system with both spin-dependent electric and ferromagnetic 
 dissipation. The ferromagnetic system is not closed, since spins are 
 transfered from the electric subsystem to the ferromagnetic 
 subsystem. However, the total system is 
 closed. The total internal entropy production allows to access 
 to the kinetic equations of the coupled system. The power dissipated is now:
 
 \begin{equation}
T \frac{d S_{i}}{dt} =  \vec{j}_{tot}^F . \vec 
{\nabla}_{\Sigma} \mu
^F - \vec{\delta J}^e . \frac {\partial{\vec{\Delta {\mu}}^e}} {e \partial 
{z}} -
J_{0}^e \frac { \partial{\mu_{0}^e}} {e \partial {z}} + \dot \Psi 
\Delta \mu^{e}
\label{source0}
\end{equation}
 
where $\vec J_{tot}^{F}$ is the total ferromagnetic flux that includes
spin transfer and where the vectorial form of the pumping force $\vec{
\Delta \mu}^{e}$ is introduced in order to take into account the
transverse spin-accumulation mechanisms discussed in the litterature
related to microscopic theories of spin-transfer-torque.

Ignoring the electric dissipation (i.e. the two last terms in Eq. 
(\ref{source0})), the
following form is obtained for the ferromagnetic system (after some crude simplifications see Section IV):

\begin{equation}
\left( \begin{array}{c}
 \vec J^{F}_{tot} \\
 \vec J_{Q}^{F} \\
\end{array} \right)  = 
 \left( \begin{array}{cc}
 \bar{\mathcal L}  &  \bar l \\
 \bar {\tilde l}  & \bar \sigma   \\
\end{array} \right) 
 \left( \begin{array}{c}
 \vec \nabla_{\Sigma} \mu^{F} \\
 \vec{\Delta \mu^{e}} \\
\end{array} \right)
\label{Result}
\end{equation}

where the matrices $\bar{\mathcal L}$, $\bar l$, $\bar {\tilde l} $
and $\bar \sigma$ are related to measurable experimental parameters. 
In the same way as for the thermoelectric power in Eq. (\ref{TEP0}), the 
presence of the
cross-coefficients is justified by the fact that
the diffusion of the spin carriers at the interface (e.g. $s-d$ 
relaxation) is contributing to the transport of 
ferromagnetic moments.

The consequences of the suplementary term in the expression of the
current $\vec J_{tot}^{F}$ in Eq.(\ref{Result}) are investigated in
terms of a generalized Landau-Lifshitz equation that includes drift and
diffusion contributions due to spin-transfer \cite{PRB08} (Section IV).  The rough
arguments presented in this introductory section will be developed and
detailed in the following sections.

\section{Derivation of Landau-Lifshitz equation from the 
corresponding power dissipation}

 In order to treat statistically the time dependence of a unique
 uniform ferromagnetic moment $\vec M = M_{s} \vec u_{r}$ (with radial
 unit vector $\vec u_{r}$) of a fluctuating magnetic nanostructure,
 the ergodic property is used.  It allows work with a statistical
 ensemble of a large number of ferromagnetic moments $\vec m$ oriented
 in the direction $\{ \theta \pm d \theta, \varphi \pm d \varphi\}$ of
 a sphere $\Sigma$ of radius $M_{s}$.  The density
 $\rho^{F}_{0}(\theta,\varphi)$, defined on the surface of the sphere, is
 then identified with the statistical distribution of ferromagnetic
 moments.  The introduction of the density is justified by the
 nanoscopic size of the magnetic single domain, for which the
 fluctuations play a major role.  To that point of view, the system is
 mesoscopic.  Accordingly \cite{Mazur}, the chemical potential
 $\mu^{F}_{0} $ takes the general form :

\begin{equation}
   \mu^{F}_{0} = kT \, ln(\rho^{F}_{0}) + V^{F}
   \label{MuF}
\end{equation}

in which the ferromagnetic potential is for instance $V^{F}(\vec H,\theta) = K
sin(\theta) - M_{s} H cos(\theta - \phi)$ in the case of a single 
domain with uniaxial anisotropy of constant $K$ and with an
external magnetic field $\vec H$ applied at an angle $\phi$ from the
anisotropy axis.

The subscript $0$ stands for a closed ferromagnetic system (no source of
magnetic moments). The corresponding current of magnetic moments 
 $\vec J_{0}^{F}$ is related to the density by the conservation law:

\begin{equation}
\frac{d\rho^{F}_{0}}{dt}= - div_{\Sigma} \vec J_{0}^{F}
   \label{conservation}
\end{equation}

where $div_{\Sigma} $ is the divergence operator defined on the surface 
of the sphere $\Sigma$.

The power dissipated by the ferromagnetic system is given by the
corresponding internal entropy production $\frac{dS^{F}_{i}}{dt}$, and
is given by the product of the generalized flux by the generalized
force.  Assuming a uniform temperature $T$ we have:
 
\begin{equation}
T \frac{dS^{F}_{i}}{dt} = 
\vec J_{0}^{F} . \vec \nabla_{\Sigma} \mu^{F}_{0}
\end{equation}

The application of the second law of thermodynamics $dS^{F}_{i}/dt \ge 0 $
allows the transport equation to be deduced
by writing the relation that links the generalized flux (the current
$\vec J_{0}^{F}$) of the extensive variables under consideration and to the generalized
force defined in the corresponding space $\Sigma$.  Both quantities, flux and
forces, are related by the Onsager matrix of the transport
coefficients $\bar{\mathcal L}_{0}$:

\begin{equation}
    \vec J_{0}^{F} = \bar{\mathcal L}_{0} \, \vec \nabla_{\Sigma} 
    \mu^{F}_{0}
   \label{Jferro}
\end{equation}

The problem is
solved as soon as the Onsager matrix is known.  In the present case,
we started from the hypothesis that the magnetic domain was uniform: the
modulus of the magnetization is conserved.  The trajectory of the
magnetization (in the configuration space) is then confined on the
surface of a sphere of radius $M_{s}$, and the flow is a two component
vector defined with the unit vectors $\{ \vec u_{\varphi}, \vec
u_{\theta} \}$ of $\Sigma$.  Accordingly, the Onsager matrix is a 2 by 2 matrix
defined by four transport coefficients $\{L_{\theta \theta},
L_{\theta \varphi}, L_{\varphi \theta}, L_{\varphi \varphi} \}$. 
Furthermore, the Onsager reciprocity relations impose that $L_{\theta
\varphi} = - L_{\varphi \theta}$.

However, the magnetization is defined by a given axis (unit vector $\{
\vec u_{r}\}$) in 3D space.  The choice of the two other vectors
is arbitrary, so that $L_{\theta \theta} = L_{\varphi \varphi}$.  Let
us now introduce a dimensionless supplementary coefficient $\alpha$,
which is the ratio of the off diagonal to the diagonal coefficients:
$\alpha = L_{\theta \varphi} / L_{\theta \theta} $.  In conclusion,
the ferromagnetic kinetic equation is defined by two ferromagnetic
transport coefficients $L_{\theta \varphi} = \rho_{0}^{F} L_{F}$ and
$\alpha$:

\begin{equation}
\bar{\mathcal L}_{0} =
\rho_{0}^{F} L_{F} \,  \left( \begin{array}{cc}
 \alpha &  1 \\
                        - 1 & \alpha \\
\end{array} \right)
\label{MatrixL0}
\end{equation}

On the other hand, the generalized force $\vec \nabla_{\Sigma} \mu^{F}_{0} $, 
thermodynamically conjugated to the magnetization, defines a 
"generalized" effective magnetic field 

\begin{equation}
\vec H_{eff} \equiv - \vec \nabla_{\Sigma} 
\mu^{F}_{0} 
\end{equation}.

It is a generalization in the sense that this effective field includes the 
diffusive term \cite{Raikher} that has first been introduced by Brown in the 
rotational Fokker-Planck equation \cite{Brown}. 

The equation Eq. (\ref{Jferro}) is the well known phenomenological 
Landau-Lifshitz (LL) equation: 

\begin{equation}
\vec J_{0}^{F} = 
- \rho_{0}^{F} L_{F} \,  \left( \begin{array}{cc}
 \alpha &  1 \\
                        - 1 & \alpha \\
\end{array} \right)
 \vec H_{eff}
    \label{BasicLL}
\end{equation}

Actually, it could be rather surprising to claim that Eq. 
(\ref{BasicLL}), that takes the form of the Fick's law or
thermoelectric
laws (with the cross-coefficients),  is the "well-known LL equation" because the LL equation
is the dynamical equation of the ferromagnetic variable $\vec M = M_{s} \vec
u_{r}$.  However, it is sufficient to rewrite Eq.  (\ref{BasicLL}) in 3D
space with re-introducing the radial unit vector $ \vec u_{r} =
(1,0,0) $ of the reference frame $\{ \vec u_{r}, \vec u_{\theta}, \vec
u_{\varphi} \}$, and recalling that the current is the density 
multiplied by
the velocity $\vec J_{0}^{F} = \rho^{F}_{0} d\vec u_{r}/dt$, in order to 
recover the traditional LL equation from (\ref{BasicLL}):

\begin{equation}
\frac{d\vec u_{r}}{dt}  =
 L_{F} \, \left \{ 
 \vec u_{r} \times \vec H_{eff} -  \alpha \vec u_{r} \times \left ( \vec u_{r} \times
 \vec H_{eff} \right \} 
 \right ) 
 \label{LL}
\end{equation}

Furthermore, it is well-known that LL equation is equivalent to the
following Gilbert \cite{Gilbert} equation, that allows the damping
coefficient $\eta$ to be defined:

\begin{equation}
\frac{d\vec u_{r}}{dt}  =
 \vec u_{r} \times \Gamma  \left (
  \vec H_{eff} - \eta M_{s} \, \frac{d\vec u_{r}}{dt}
 \right ) 
\end{equation}

where $\Gamma$ is the gyromagnetic ratio.
The equivalence between the two equations defines the 
coefficients $\alpha$ and $L_{F}$ has a function of the coefficients $\eta$ 
and $\Gamma$. $\alpha$ is the dimentionless damping coefficent:

\begin{equation}
    \alpha = \eta \Gamma M_{s}
    \label{alpha}
\end{equation}

and $L_{F}$ is defined by the relation

\begin{equation}
         L_{F}
		 =  \frac{\Gamma}{M_{s} \left (1+ \alpha^{2} \right )}
\label{Ltheta}
\end{equation}
 
The corresponding Fokker-Planck stochastic equation first derived by Brown \cite{Brown} is 
obtained directly by inserting 
Eq. (\ref{BasicLL}) into Eq.(\ref{conservation}).

\begin{equation}
\frac{d\rho^{F}_{0}}{dt}= - \vec \nabla_{\Sigma} \bar{\mathcal L}_{0}  \vec \nabla_{\Sigma} 
    \mu^{F}_{0}
   \label{FPEMag}
\end{equation}

 Using Eq. (\ref{MuF}), Eq. (\ref{alpha}), Eq. (\ref{Ltheta}), 
and the explicit expression of the Laplacian $\nabla^{2}_{\Sigma}$ in spherical 
coordinates, the Fokker-Planck equation reads:
 
\begin{equation}
   \begin{aligned}
    \frac{\partial \rho_{0}^{F}}{\partial t}
     = & 
     \frac{L_{F}}{\sin\theta} 
     \frac{\partial}{\partial \theta}
     \left\{
       \sin\theta
        \left[
          \left(
            \alpha \frac{\partial V^{F}}{\partial \theta}
          -
            \frac{1}{\sin\theta}
            \frac{\partial V^{F}}{\partial \phi}
          \right)
          \rho_{0}^{F}
        +
          kT \alpha \frac{\partial \rho_{0}^{F}}{\partial \theta}
        \right]
     \right\} \\
    &\quad
    +
     \frac{1}{\sin\theta}
     \frac{\partial}{\partial \phi}
     \left\{
       \left(
        \frac{\partial V^{F}}{\partial \theta}
      +
        \frac{\alpha}{\sin\theta}
        \frac{\partial V}{\partial \phi}  
       \right)
       \rho_{0}^{F}
     +
       kT 
       \frac{\alpha}{\sin\theta}
       \frac{\partial \rho_{0}^{F}}{\partial \phi}
     \right\} 
   \end{aligned} 
   \label{FPE2}
  \end{equation}
  
The driving force responsible for the magnetization dynamics is
distributed between drift and diffusion terms for the probability
distribution.  The equilibrium solution of the equation is the
Boltzmann distribution, as it was assumed in expression Eq.  (\ref{MuF}) for
the definition the chemical potential .

Since the equation depends on the determinist potential $V^{F}$ (that
contains the energy due to the external magnetic field, the
magnetocrystaline anisotropy, dipolar energy, etc), it is non linear. 
Only a few simple configurations can find an analytical solution for the
non-equilibrium statistical distribution $\rho_{0}^{F}(\theta,
\varphi,t)$ \cite{Coffey}.  This is typically the case for linear
expansions near equilibrium states in the context of ferromagnetic
resonnance, or for the N\'eel-Brown activation process 
\cite{Brown,Coffey} 
at long time scales. 
Eq.  (\ref{FPE2}) will be extended to spin-transfer contributions
introduced in Section III, after the study of spin-accumulation
effects below.

\section{Derivation of spin-accumulation from the 
corresponding power dissipation}

In this section, we focus on the electric transport only (we forget
the role of the ferromagnetic variable and the existence of the
$\Sigma$ space).  The corresponding electric wire is defined along the
$z$ axis, with a section unity: the relevant configuration space is
the one dimentional real space $\Re$.  However, in order to take into
account the spin-dependent electric current, the two channel model is
introduced. Beyond the diffusive two chanel model approach, transverse spin 
accumulation parameters are also introduced, in order to take into account the 
spin precession in quasi-ballistic regime near the 
interface.

\subsection{two channel relaxation and spin-diffusion}

The system is composed of two populations of conduction electrons with
a relaxation process that allows the electrons of one population relax
into the other.  The difference between the two populations is
introduced through an internal degree of freedom, and the relaxation
process occurres within the space defined by this internal degree of
freedom.  Inside the bulk, the relaxation from one channel to the
other is compensated by the opposit relaxation: the electronic
populations are maintained at equilibrium.  However, the presence of
an interface with inhomogeneous transport parameters puts
out-of-equilibrium the electronic populations.  At steady state, a
diffusion process (of the spin-density) in the $\Re$ space occurres that compensates the
forced relaxation defined in the internal space.  This diffusion
process is called spin-accumulation in the case of spin-dependent
transport.  Formally, the two-channel model consisted of defining a
bi-valuated internal variable for the transport parameters, that takes
the values $\alpha$ or $\gamma$.  Typically, the values of the
internal variables are $\alpha = \uparrow$ and $\gamma = \downarrow$
for spin-flip scattering, or $\alpha = s$ ($\uparrow \, \downarrow $)
and $\gamma = d$ ($\downarrow$) for spin-dependent $s-d$ scattering
\cite{Revue,MTEP} ( the three channel model in which the internal
variable takes the values $ \{ s \uparrow, s \downarrow, d \downarrow
$) is developed in reference \cite{Revue}).  Accordingly, the local
electrochemical potentials are defined by $\mu^{e}_{\alpha}$ and
$\mu^{e}_{\gamma}$, and the electric currents generated in each
channel is noted $ \{J^{e}_{\alpha}, J^{e}_{\gamma}\}$
  
The conservation laws write: 

\begin{equation}
\left\{
        \begin{array}{c}
\frac{d n_{\alpha}}{dt}\, = \,-\frac{\partial J^{e}_{\alpha}}{\partial
z} - \, \dot{\Psi} \\
\frac{d n_{\gamma}}{dt}\, = \,-\frac{\partial J^{e}_{\gamma}}{\partial
z}
+ \, \dot{\Psi} \\
\end{array}
\label{con0}
\right.
\end{equation}

where $n_{\alpha}$ and $n_{\gamma}$ are the densities of charge
carriers in the channels $\{\alpha, \gamma \}$, and the spin-dependent
relaxation is taken into account by the flux $\dot \Psi $.  This is
the velocity of the reaction (or relaxation of the spin-dependent
internal variable) that transforms a conduction electron $\alpha $
into the conduction electron $\gamma $.  This generalized flux defines
a "spin current" (density times velocity) in the configuration space
of the internal variable (somehow related to $\Sigma$: see next
section).  Note however that in the litterature the term "spin current" is
devoted to the spin-polarized electric current $\delta J^{e} =
J^{e}_{\alpha} - J^{e}_{\gamma}$ defined in the real space $\Re$.

The power dissipated by the electric system is given by the corresponding internal entropy 
variation, i.e. by the product of the currents by the electric fields:

\begin{equation}
T \frac{dS^{e}_{i}}{dt} = -J^{e}_{\alpha} . \frac{\partial 
\mu^{e}_{\alpha}}{e \partial
z} - J^{e}_{\gamma} . \frac{\partial \mu^{e}_{\gamma}}{e \partial z} + \dot
\Psi . \Delta \mu^{e}
\label{entropyTC}
\end{equation}

where we introduced the difference of the chemical potentials $\Delta
\mu^{e} = \mu^{e}_{\alpha} - \mu^{e}_{\gamma}$ \cite{Wyder,ValetFert}.

the application of the second law of thermodynamics leads to the 
kinetic
equations, after introducing the transport coefficients: the
conductivities $\sigma_{\alpha}$, $\sigma_{\gamma}$ , and the Onsager
coefficient $L $, such that:

\begin{equation}
\left\{
\begin{array}{c}
J^{e}_{\alpha} = -\frac{\sigma_{\alpha }}{e} \frac{\partial
\mu_{\alpha}}{\partial z}\\
J^{e}_{\gamma} = -\frac{\sigma_{\gamma}}{e} \frac{\partial
\mu_{\gamma}}{\partial z}\\
\dot{\Psi} = L \Delta \mu^{e}
\end{array}
\right. \label{OnsagerElec0}
\end{equation}

where the two first equations are Ohm's law applied to each
channels.  The effect of the electric charge relaxation is
described in reference \cite{Revue}.  The
Onsager coefficient $L$ is shown to be inversely proportional to the
electronic relaxation times $\tau_{\alpha \leftrightarrow \gamma}$. 
The total electric current is spin-independent:

\begin{equation}
J^{e}_{0} = J^{e}_{\alpha} + J^{e}_{\gamma} = -\frac{1}{e} \frac{\partial
}{\partial z} \left (\sigma_{\alpha } \mu^{e}_{\alpha}+ \sigma_{\gamma }
\mu^{e}_{\gamma } \right )
\end{equation}

However, it is not possible to measure separately the different
conduction channels, since any realistic electric contact short cuts
the two channels.  What is measured is necessarily the usual Ohm's
law, $J_{0}^{e}= -2 \sigma_{0} \frac{\partial \zeta}{\partial z}$, that
imposes the reference electric potential $\zeta$ to be introduced,
together with the mean conductivity $\sigma_{0}= (\sigma_{\alpha}+
\sigma_{\gamma})/2$. The potential $\zeta$ is hence:

\begin{equation}
e \zeta= \frac{2}{\sigma_{0}}( \sigma_{\alpha} \mu^{e}_{\alpha} +
\sigma_{\gamma} \mu^{e}_{\gamma})
\end{equation}

The reference configuration is defined by the two channels collapsing
to a unique conduction channel (e.g. parallel magnetization of a
junction of two identical ferromagnetic layers: $\Delta
\mu^{e}_{eq}(0)=0$ ).  The non-equilibrium ($\Delta
\mu^{e}(0)\ne 0$) contribution of the junction to the resistance, $R^{ne}$, is
calculated through the relation:

\begin{equation}
J_{0}^{e}e \, R^{ne}  = \int_{A}^{B}  \frac{\partial }{ \partial z}
(\mu^{e}_{\alpha} - e \zeta (z))dz = \int_{A}^{B}  \frac{\partial }{
\partial z} (\mu^{e}_{\gamma} - e \zeta(z))dz \label{Integ}
\end{equation}

so that

\begin{equation}
      R^{ne}= -\frac{1}{J_{0}^{e}e} \int_{A}^{B}
\frac{\sigma_{\alpha} - \sigma_{\gamma}}{\sigma_{0}}
\frac{\partial \Delta \mu^{e}}{ \partial z}dz 
\label{Rneq0}
\end{equation}

where the measurement points $A$ and $B$ are located far enough in
each side of the interface (inside the bulk) so that $\Delta \mu^{e}
(A)=\Delta \mu^{e} (B) =0$.  The integral in Eqs.  (\ref{Integ}) is performed over
the regular part of the function only (across the interface $\zeta$
and $\sigma_{i}$ are discontinuous at this scale): this resistance is
proportional to the discontinuity at the interface.  It is convenient
to describe the conductivity asymmetry by the parameter $\beta$ such
that $\sigma_{\alpha}= \sigma_{0} (1+\beta)$ and
$\sigma_{\gamma}=\sigma_{0}(1-\beta)$.  On the other hand, the
spin-polarized electric current is $\delta J^{e} = J^{e}_{\alpha} -
J^{e}_{\gamma} $.  With these new variables, Eq.(\ref{Rneq0}) rewrites:

\begin{equation}
      R^{ne}= -\frac{2 \beta}{J_{0}^{e}e} \int_{A}^{B}
\frac{\partial \Delta \mu^{e}}{ \partial z}dz 
\label{Rneq}
\end{equation}

and the Onsager matrix reads:

\begin{equation}
\left( \begin{array}{c}
\delta J^e \\
J^{e}_{0} \\
\dot \Psi 
\end{array} \right)
=  \left( \begin{array}{ccc}
                        \sigma_0 & \beta \sigma_0 & 0\\
                       \beta \sigma_0 & \sigma_0 & 0 \\
			 0 & 0 & L
                        \end{array} \right)
                        \left( \begin{array}{c}
\frac{-1}{e} \frac {\partial{\Delta 
{\mu}^{e}}}{\partial {z}}\\
\frac{-1}{e} \frac{\partial \mu^{e}_{0}}{\partial z}\\
\Delta \mu \\
 \end{array} \right)
\label{OnsagerElec1}
\end{equation}

The system of equations Eq.  (\ref{OnsagerElec1}) allows the diffusion
equation for $\Delta \mu(z)$ to be derived for the stationary 
conditions $ \partial \vec J_{0}^{e}/ \partial z = 0$ and $\partial 
\vec{\delta J}^{e}/ \partial z = - 2 \dot \Psi$:

\begin{equation}
\frac{ \partial^{2} \Delta \mu}{\partial z^{2}} = \frac{\Delta 
\mu}{l^{2}_{diff}} 
\label{Diff}
 \end{equation}
 
 where $l_{diff}^{-2} = \frac{2eL}{\sigma_{0}(1-\beta^{2})}$. The resistance 
 $R^{ne}$ can then be calculated for each specific
 device configurations \cite{Revue,MTEP}.

The two channel approximation with the internal degree of freedom that
takes the value $\{\uparrow, \downarrow \}$ describes the consequences
of the spin-flip scattering. However, the model is also sufficient for the
description of spin-dependent $s-d$ relaxation \cite{Revue,MTEP}, where 
the $d$ band is full for the majority spins $\uparrow$. Indeed, 
the $s-d$ relaxation with spin-flip (from $s \uparrow$ to $d 
\downarrow$) has a very small probability to occurre, compared to the relaxation 
without spin-flip (from $s \downarrow$ to $d 
\downarrow$). As a consequence, the two channel model can also be 
used, and it also leads to a redistribution of the spin populations at 
the interfaces, i.e. to spin-accumulation.  In
both cases, the spin-accumulation is described by the function $\Delta 
\mu(z) $, which is solution of is the the diffusive equation Eq. 
(\ref{Diff}).

\subsection{Quasi-ballistic effect and transverse spin-accumulation}

However, the description proposed above with a spin-dependent internal variable that takes
the two values $\{\alpha, \gamma\}$ is not able to take
into account the precession of the spins occuring in a magnetic field. 
If the precession contribution is
not relevant in the case of the processes
that lead to giant magnetoresistance (because the mean values are
averaged out over the spin-diffusion length) this is no longer the case
in a quasi ballistic regime near the interface.

In order to take into account quasi-ballistic effects near the interface 
(i.e. sub-nanometric scales in metalic devices),
the two-channel model has been recently generalized to transverse
spin-accumulation in the context of spin-transfer-torque
investigations \cite{LevyFert,Levy1,Levy2,Levy3,Dugaev,BarnasFert}. 
The transverse spin-accumulation is introduced with the corresponding
current $\delta J^e_{\perp}$ and the corresponding chemical potential
$\Delta {\mu}^{e}_{\perp}$. Transverse means here that the spin 
density is considered in the plan perpendicular to the 
quantification axis $\updownarrow$ that defines the spin up and spin 
down in the two channel-model.

The coefficient $\sigma_{\perp}$ can also be defined through the
corresponding diffusion coefficient $D_{\perp} =
\frac{\sigma_{\perp}kT}{n_{\perp}}$ where $n_{\perp}$ is the density
of transverse spins \cite{Chung}.  It is then also possible to define
a "pseudo" spin-diffusion process in the case of spin-decoherence.  Note however
that the two potentials $\Delta \mu^{e}$ and $\Delta \mu_{\perp}^{e}$
are defined at very different length scales and it is necessary to
refer to quantum approaches in order to understand the physical
signification of the transverse parameters \cite{Waintal,Braatas}. 
The corresponding contribution to the power dissipated is

\begin{equation}
T \frac{dS^{e}_{\perp}}{dt} = - \delta J^{e}_{\perp} . \frac{\partial 
\Delta \mu^{e}_{\perp}}{e \partial
z}
\label{entropy}
\end{equation}

Puting all together we have the following Onsager relations for the electric 
system:

\begin{equation}
\left( \begin{array}{c}
\delta J^e \\
J^{e}_{0} \\
\delta J^{e}_{\perp}\\
\dot \Psi 
\end{array} \right)
=  \left( \begin{array}{cccc}
                       \sigma_0 & \beta \sigma_0 & 0 & 0 \\
                       \beta \sigma_0 & \sigma_0 & 0 & 0 \\
		        0  &   0  &  \sigma_{\perp} &  0 \\
			 0 & 0 & 0 & L
                        \end{array} \right)
                        \left( \begin{array}{c}			
\frac{-1}{e} \frac {\partial{\Delta 
{\mu}^{e}}}{\partial {z}}\\
\frac{-1}{e} \frac{\partial \mu^{e}_{0}}{\partial z}\\
\frac{-1}{e} \frac {\partial{\Delta 
{\mu}^{e}_{\perp}}}{\partial {z}}\\
\Delta \mu \\
 \end{array} \right)
\label{OnsagerElec2}
\end{equation}

\section{Derivation of Spin-transfer due to longitudinal and transverse 
spin-accumulation}

In usual experimental configurations for spin-transfer, an electric
current is injected in a circuit that includes a ferromagnet (in
series or in non-local configuration
\cite{Otani}) and the magnetoresistance, i.e. the potential drop (of
the form $- \frac{\beta}{2} \int_{A}^{B}\frac{\partial \Delta \mu}{e
\partial z}dz$ Eq.  (\ref{Rneq})) allows the
magnetization states to be measured. The effect of strong electric 
currents on the magnetization
states can then be observed.  In such a configuration, the two sub-systems
described in the sections above exchange magnetic moments at the
junctions and both are open systems.

In order to describe the dynamics of the ferromagnetic degrees of
freedom (following step by step the method presented in Section II), we have to
deal with a closed system.  The system of interest is now the
ferromagnetic system that includes spin-accumulation effects at the
junctions. 
This total ferromagnetic system is such that the density of 
ferromagnetic moments $\rho^{F}_{tot}$ and the total ferromagnetic flux $\vec 
J^{F}_{tot}$ are related by the conservation law:  $d \rho^{F}_{tot}/dt = -
div_{\Sigma} \vec J^{F}_{tot}$.

The initial
configuration space of magnetic moments is then extended
to 1D real space parametrized by the internal
variable $\Sigma \otimes \Re_{\alpha \gamma}$.  The
important point here is that the internal variable is spin-dependent, and
related to the ferromagnetic space $\Sigma$ (e.g. through $s-d$ 
relaxation and the corresponding spin-accumulation). This accounts for the 
coupling, i.e. the {\it transfer}, of magnetic moment between the two 
sub-systems.

The dissipation is given by the internal power dissipated in the total system 
$ T \, d S_{i}/dt$ :

\begin{equation}
T \, \frac{d S_{i}}{dt} = \vec{j}_{tot}^F . \vec 
{\nabla}_{\Sigma} \mu
^F - \delta J^e . \frac {\partial{\Delta {\mu}^e}} {e \partial 
{z}} - \delta J^e_{\perp} . \frac {\partial{\Delta {\mu}^e_{\perp}}} {e \partial 
{z}} -
J_{0}^e \frac { \partial{\mu_{0}^e}} {e \partial {z}} + \dot \Psi 
\Delta \mu^{e}
\label{source}
\end{equation}

Where the first term in the right hand side is the power dissipated by 
the total ferromagnetic sub-system (including the ferromagnetic contribution due 
to spin-transfer), the two following terms are the 
power dissipated by spin-dependent electric transport, and the fourth term 
is the spin-independent Joule 
heating. The last term is the power dissipated by spin-flip
or $s-d$ relaxation.

In Eq.  (\ref{source}), the vectors are defined on the
sphere $\Sigma$ with the help of two angles $\theta$ and $\varphi$. 
The total ferromagnetic current $\vec{j}_{tot}^F = j_{tot}^{F \theta}
\vec u_{\theta} + j_{tot}^{F \varphi} \vec u_{\varphi} $ includes the
contribution due to spin-accumulation mechanisms. The chemical potential
$\mu^{F}$ accounts for the energy of a ferromagnetic layer. On the
other hand, the system is contacted to electric reservoirs with the 
electric currents and the corresponding chemical potentials.
Applying the second law of thermodynamics, we obtain the general Onsager 
relations:

\begin{equation}
\left( \begin{array}{c}
j_{tot}^{F \varphi} \\
j_{tot}^{F \theta} \\
\delta J^e_{\perp} \\
\delta J^e \\
J^{e}_{0} \\
\dot \Psi 
\end{array} \right)
=   \left( \begin{array}{cccccc}
                        \alpha \rho_{0} L_{F} & \rho_{0} L_{F} & 
                        l_{\varphi \varphi} & l_{\varphi \theta} & 0 & 0\\
                        - \rho_{0} L_{F} & \alpha \rho_{0} L_{F} &  l_{\theta \varphi} & 
                        l_{\theta \theta} & 0 & 0\\
			\tilde l_{\varphi \varphi} & \tilde l_{\varphi \theta} & \sigma_{\perp} & 0  & 0 & 0 \\
                        \tilde l_{\theta \varphi} & \tilde l_{\theta \theta} & 0 & \sigma_0 & \beta \sigma_0  & 0\\
                        0 & 0 & 0 & \beta \sigma_0 & \sigma_0 &  0 \\
						0 & 0 & 0 & 0 & 0 & L
                        \end{array} \right)
                        \left( \begin{array}{c}
 \frac{1}{sin( \theta)} \frac{\partial \mu ^F}{\partial \varphi}\\
 \frac{\partial \mu ^F}{\partial \theta}\\
 \frac{-1}{e} \frac {\partial{\Delta 
{\mu}^{e}_{\perp}}}{\partial {z}} \\
\frac{-1}{e} \frac {\partial{\Delta 
{\mu}^{e}}}{\partial {z}}\\
\frac{-1}{e} \frac{\partial \mu^{e}_{0}}{\partial z}\\
\Delta \mu \\
 \end{array} \right)
\label{Onsagermatrix}
\end{equation}

All coefficients were defined in the previous sections, except
the new cross-coefficients $ \{ l_{\varphi \varphi}, l_{\varphi
\theta}, l_{\theta \theta}, l_{\theta \varphi} \}$, introduced in this
model as spin-transfer coefficients, related to the
experimental parameters.  The coefficients $\{\tilde l_{i}\}_{i=\{ 
\theta, \varphi\}}$ are
deduced from the coefficients $ \{ l_{i}\}$ through the Onsager
reciprocity relations. 

The total ferromagnetic current can be written after integrating over the volume 
$v$ 
of the ferromagnetic layer of section unity and the spin accumulation 
zone. This volume is such that $v = 
\int_{A}^{B} dz$, where $z = A$ and $z = B$ are two sections close to the interface but 
far enough with
respect to the diffusion length $l_{diff}$. We assume 
here that $l_{diff}$ 
is much smaller than the width of the ferromagnetic layer in order to 
simplify the calculation: the volume of the ferromagnetet is 
identified as $v$.
Let us define $ \vec X$ as
the correction due to the spin-transfer deduced from 
the two first equations of the matrix equation Eq. 
(\ref{Onsagermatrix}), after integrating over the volume $v$ :

\begin{equation}
   v  \vec J_{tot}^{F} = v \bar{\mathcal L}_{0} \, \vec \nabla \mu^{F}_{0} + 
   \vec X 
   \label{JtotST}
\end{equation}

where $\bar{\mathcal L}_{0}$ is the matrix defined in Eq. (\ref{MatrixL0}).

The assumption of constant modulus of the magnetization imposes 
that $\vec X $ is confined on the surface of the sphere $\Sigma$. The 
Helmoltz decomposition theorem can then be applied: the vector $\vec X$  can be 
decomposed in a unique way with the introduction of the two potentials $\chi$ and $\Phi$ 
(i.e. a 
potential vector) such that \cite{Serpico}:

\begin{equation}
\vec X = \vec u_{r} \times \vec \nabla_{\Sigma} \Phi + \vec \nabla_{\Sigma} \chi
\label{Helmoltz}
\end{equation}

where the first term is divergenceless and the second term is 
curless (i. e. non conservative). The method used here is hence not 
equivalent to that of adding a spin-transfer source term in the conservation 
equation of the time variation of $\rho^{F}_{tot}$. The two potentials 
will be described in more details below.

The total correction to the Landau-Lifshitz-Gilbert equation 
writes:

\begin{equation}
    v \vec J_{tot}^{F} = v \bar{\mathcal L}_{0} \vec \nabla_{\Sigma} \mu_{t} +  
    \vec u_{r} \times \vec \nabla_{\Sigma} \Phi
    \label{Correc}
\end{equation}

where the electro-spin chemical potential that describes the 
ferromagnetic system with the addition of the spin-transfer 
contribution writes 
\begin{equation}
\mu_{t} = kT \, ln(\rho^{F}_{0}) + V^{F} + \frac{\chi}{L_{F}} 
\label{mutot}
\end{equation}

However, in Eq. (\ref{mutot}), the density $\rho^{F}_{0}$ is no longer relevant because the ferromagnetic
system alone is an open system, and only the total density $\rho^{F}_{tot}$
is defined. The canonical form of the
chemical potential of the total system that contains the total density $\rho_{tot}$ is:
$\mu^{F}_{t} = kTln(\rho^{F}_{tot}) + V^{F}$. 
The total density is deduced with identifying with Eq. 
(\ref{mutot}) 
$\rho^{F}_{tot} =\rho^{F}_{0}e^{\frac{\chi}{kT L_{F}} }$

The generalized LLG takes the form:

\begin{equation}
\frac{d\vec{u}_{r}}{dt} =\frac{\gamma}{M_{s}(1 + \tilde{\alpha}^{2})} \left \{
\vec{u}_{r}\times \left (\vec{H}_{eff} +
\frac{\vec \nabla_{\Sigma} \Phi}{v \rho^{F}_{0} L_{F}} \right) \\
-  \alpha \vec{u}_{r} \times \left[
    \vec{u}_{r}\times \left ( \vec{H}_{eff} - \frac{\vec \nabla_{\Sigma} \chi}{v \rho^{F}_{0} L_{F}  \alpha}  \right ) \right]
\right \}
\end{equation}

{\it This is the main result of this work}.

The corresponding Fokker-Planck equation is obtained by inserting the
expression of $\vec J_{tot}$ into the conservation equation: $d
\rho^{F}_{tot}/dt = - div_{\Sigma} \vec J^{F}_{tot}$. The study of 
the resulting stochastic equation is however beyond the aim of this 
paper.

In order to give an expression of the two potential-energy terms \{$\chi$, 
$\Phi$\}, we
will make the following assumption.  According to previous discussions
\cite{MTEP,PRB03,PRB08} based on the separation of the typical
relaxation time scales involved during the ferromagnetic processes,
the usual spin-accumulation due to spin-dependent relaxation $\Delta
\mu^{e} $ is coupled to the relaxation of the magnetization
$j_{tot}^{\theta F}$ only (because $l_{diff}$ defines a mesoscopic
variable that scales with the magnetization).  On the other hand, we assume that the transverse spin
accumulation is coupled to the precession only (i.e. acting at subnanosecond time-scale).

In this case and after integrating over the 
volume $v$ the ferromagnetic current writes :

\begin{equation}
\left\{
\begin{array}{c}
v J_{tot}^{F \theta} =  v J_{0}^{F \theta} + \frac{\partial 
\chi}{\partial \theta} - \frac{1}{sin \theta}\frac{\partial 
\Phi}{\partial \varphi}\\
v J_{tot}^{F \varphi} =  v J_{0}^{F \varphi} + \frac{\partial 
\Phi}{\partial \theta}
\end{array}
\right. 
\end{equation}

On one hand the potential $\chi$ is directly associated to the
voltage drop due to spin-dependent relaxation (Eq. \ref{Rneq}):

\begin{equation}
\frac{\partial \chi}{\partial \theta } =  \frac{-l_{\theta \theta} }{e} 
\int_{A}^{B}\frac{\partial \Delta \mu^{e}}{\partial z} dz = 
-\frac{J_{0}^{e} \, l_{\theta \theta}}{2 \beta} \, R^{GMR}
\label{Vdrop}
\end{equation}

and this expression can be generalized to specific device configurations. 
The potential energy $\chi$, function of the magnetic coordinates, 
can be measured in the context of two-level-fluctuation experiments performed on 
individual magnetic nanostructures \cite{PRB03,PRB08}.

On the other hand, the potential $\Phi$ is associated to the discontinuity of the transverse spin 
accumulation and is responsible for the spin-transfer torque:

\begin{equation}
\left\{
\begin{array}{c}
\frac{\partial 
\Phi}{\partial \varphi} =  \frac{l_{\varphi \theta} sin \theta}{e} \int_{A}^{B} \frac{\partial 
\Delta \mu_{\perp}}{\partial z} dz \\
  \frac{\partial \Phi}{\partial \theta} =  - \frac{l_{\varphi \varphi}}{e} \int_{A}^{B} \frac{\partial 
\Delta \mu_{\perp}}{\partial z} dz
\end{array}
\right. 
\end{equation}

In these expressions, and in analogy with 
spin-accumulation due to the spin-dependent relaxation, the voltage drop $\int_{A}^{B} \frac{\partial
\Delta \mu_{\perp}}{e \partial z} dz = J_{0}^{e} R_{eff}^{trans}$ is also able to define a 
non-equilibrium interface magnetoresistance $R_{eff}^{trans}$ for transverse 
spin-accumulation. Experimentally, the potential $\Phi$ is identified 
to the so called Slonczewski term used in the context of resonance 
experiments (FMR) performed 
in the GHz range.

\section{Conclusion}

In order to describe spin-transfer effect, spin-accumulation has been
taken into account explicitly in the dynamical equation of the
macroscopic ferromagnetic degrees of freedom.  This dissipative coupling
was described in terms of Onsager cross-coefficients $l_{i}$ appearing in
the Onsager matrix in Eq.  (\ref{Onsagermatrix}).

All the other terms appearing in the equation Eq. 
(\ref{Onsagermatrix}) (the other transport coefficients of the matrix,
the generalized flux, and the conjugated generalized forces) have
first been defined independently in the two preceding sections.  In
the case of the well-known dynamics of the ferromagnetic order
parameter (supposed uniform), the approach proposed allows us to define
the transport coefficients, the ferromagnetic current and the
ferromagnetic generalized force from the expression of the entropy
production and the conservation equations.  The dynamics of the
magnetization is summarized in the Onsager equation Eq. 
(\ref{BasicLL}), which is the simplest form of the well-known
Landau-Lifshitz equation.  On the other, the
spin-dependent electronic relaxations (spin-dependent $s-d$ relaxation
or spin-flip relaxation) were treated on an equal footing in the
context of the two channel model of electric conductivity.  The 
resulting
kinetic equations are also
summarized by a Onsager equation Eq(\ref{OnsagerElec1}) with the
relevant flux and forces, which are also a simple form of well known
kinetic equations (e.g. that derived from the Valet-Fert model).  The
corresponding transport coefficients and forces ($\beta$ ,
$\sigma_{0}$, $\delta J^e$, $\Delta \mu$) can be measured through the
giant magnetoresistance and related effects.  Due to quasi-ballistic 
precession of the spins of the conduction electrons, it is furthermore necessary
to generalize the spin-accumulation effect to "transverse
spin-accumulation" (according to recent reports on spin-transfer
torque).

Due to the two forms of spin-accumulation mechanisms, there are also
two forms of coupling, namelly spin-accumulation coupling (due to
spin-dependent relaxations) and transverse spin-transfer-torque
coupling due to quasi-ballistic spin precession.  The model proposes a
method able to formalize this coupling, and to deduce the consequences
in terms of Landau-Lifshitz equation.  In both cases, the coupling
between the spin of the conduction electrons and the ferromagnetic
parameters are introduce through the four phenomenological Onsager
cross-coefficients $l_{ij}$.  The generalization of the
Landau-Lifshitz equation to these contributions is performed with two
measurable potentials \{$\Psi$,$\chi$\} (functions of the magnetic
coordinates) defined in Eqs.  (\ref{Correc}) and (\ref{mutot} ).  The
potential $\chi$ is associated to the spin-accumulation generated by
the spin-dependent relaxation, and the potential $\Phi$ is associated
to the conservation of the transverse moments (spin-transfer-torque). 
The two functions \{$\Psi$,$\chi$\} are experimentally accessible.


\begin{thebibliography}{10}
     
    

\bibitem{DeGroot} S. R. De Groot and P. Mazur, {\it
Non-equilibrium thermodynamics} Amsterdam : North-Holland, 1962.

\bibitem{Prigogine} I. Prigogine, {\it Introduction to thermodynamics 
of irreversible processes}, J. Wiley and Sons, Inc., New York, 1962.

\bibitem{Stueckelberg} E.C.G.
Stueckelberg and P.B. Scheurer, {\it thermocin\'etique
ph\'enom\'enologique galil\'eenne} Birkauser Verlag, Basel and
Stuttgart, 1974.


\bibitem{Smith} A. C. Smith, J. F. Janak, R. B. Adler, {\it Electric 
conduction in solids}, McGraw-Hill Inc 1967, Chapter 1 and Chapter 2.

\bibitem{Parrott} J. E. Parrott, {\it
Thermodynamic theory of transport processes in semiconductors}, IEEE Trans electron devices 
{\bf 43}, 
809 (1996).

\bibitem{Vilar} J. M. G. Vilar and J. M. Rubi , {\it Thermodynamics ÒbeyondÓ 
local equilibrium}, PNAS {\bf 98}, 11081 (2001).

\bibitem{Regera} D. Regera, J. M. G. Vilar and J. M. Rubi, {\it The 
mesoscopic Dynamics of thermodynamic systems} J. Phys. 
Chem. B, {\bf 109}, 21502 (2005).

 \bibitem{Berger}  L. Berger, {\it Emission of spin waves by a 
magnetic multilayer traversed by a current}, Phys. Rev. B {\bf 54}, 9353 (1996).

\bibitem{Slon} J. C. Slonczewski, 
{ \it Current-driven excitation of magnetic multilayers}
J. Magn. Magn. Mat. {\bf 159} L1 (1996).

\bibitem{Hickey} M. C. Hickey and J. S. Moodera, {\it Origin of the  
Intrinsic Gilbert Damping}, Phys. Rev. Lett. {\bf 102}, 1376001 (2009).

\bibitem{Damping} M. Faehnle, R. Singer, and D. Steiauf, {\it Role of 
nonequilibrium conduction electrons on the magnetization dynamics of 
ferromagnets in the s-d model}, Phys. Rev. B {\bf 73}, 172408 (2006). 

 \bibitem{Johnson} M. Johnson and R. H. Silsbee, {\it Interfacial 
charge-spin coupling: injection and detection of spin magnetization in 
metal}, Phys. Rev. Lett. {\bf 55}, 1790 (1985) 

\bibitem{Wyder} P. C. van Son, H. van Kempen, and P. Wyder, {\it 
Boundary resistance of the ferromagnetic-nonferromagnetic metal 
interface}, Phys. Rev
Lett. {\bf 58}, 2271 (1987).

\bibitem{ValetFert} T. Valet and A. Fert,{\it Theory of the perpendicular magnetoresistance 
in magnetic multilayers},  Phys. Rev. B, {\bf 48}, 7099
(1993).

\bibitem{PRB00} J. -E. Wegrowe, {\it Thermokinetic approach of the 
generalized Landau-Lifshitz-Gilbert equation with spin-polarized 
current"}, Phys.  Rev.  B {\bf 62}, 1067
(2000).

 \bibitem{Revue} J.-E. Wegrowe, M. C. Ciornei, H.-J. Drouhin, 
{\it Spin transfer in an open ferromagnetic layer: 
from negative damping to effective temperature}
J. Phys.:
Condens. Matter {\bf 19}, 165213 (2007).

\bibitem{Mott} N. F. Mott and H. Jones, {\it Theory of the Properties 
of Metal and Alloys}, Oxford University Press, 1953.

\bibitem{Ansermet} J.-Ph. Ansermet {\it Thermokinetic 
description of spin mixing in spin dependent transport}, IEEE Trans. Mag. {\bf 44}, 329 (2008).

\bibitem{Sakurai} J. Sakurai, M. Horie, S. Araki, H. Yamamoto, and T. 
Shinjo, {\it Magnetic-field effects on thermopower of Fe/Cr and 
Cu/Co/Cu/Ni(Fe) multilayers} J. Phys. Soc. Jpn. {\bf 60}, 2522 (1991).

\bibitem{Piraux} L. Piraux, A. Fert, P. A. Schroeder, R. Laloee, and 
P. Etienne, {\it Large magnetothermoelectric power in Co/Cu, Fe/Cu, 
and Fe/Cr multilayers}, J. Magn. Magn. Mat. {\bf 110}, L247 (1992).

\bibitem{Shi} J. Shi, S. S. P. Parkin, L. Xing, 
M. B. Salamon, {\it Magnetothermopower of Co/Cu multilayers}, J. Appl. Phys. {\bf 73}, 5524 (1993).

\bibitem{Shi2} J. Shi, K. Pettit, E. Kita, S. S. P. Parkin, R. Nakatani, 
M. B. Salamon, {\it Field-dependent thermoelectric power and thermal 
conductivity in multilayered and granular giant magnetoresistive 
systems}, Phys. Rev. B,{\bf 54}, 15273 (1996).

\bibitem{Fukushima} A. Fukushima, H. Kubota, A. Yamamoto, Y. Suzuki,
{\it Peltier cooling in current-perpendicular-to-the-plane metallic 
junctions}, J. Appl. Phys. {\bf 99}, 08H706 (2006).

\bibitem{Gravier} L.
Gravier, S. Serrano-Guisan, and J. -Ph.  Ansermet, 
{\it " Spin-dependent Peltier effect in Co/Cu multilayer nanowires"} J. Appl.  Phys. 
{\bf 97}, 10C501 (2005).

\bibitem{Santi} S. Serrano-Guisan, L. Gravier, M. Abid, and
J. -Ph.  Ansermet, 
{\it Thermoelectrical study of ferromagnetic nanowire structures }, J. Appl. Phys. {\bf 99}, 08T108 (2006).


\bibitem{MTEP} J.-E. Wegrowe, Q. Anh Nguyen, M. Al-Barki, J.-F. Dayen,
T. L. Wade, and H.-J. Drouhin, {\it Anisotropic magnetothermopower: Contribution of 
interband relaxation} Phys. Rev. B {\bf 73} 134422, (2006).

\bibitem{SpSeebeck} K. Uchida, S. Takahashi, K. Harii, J. Ieda, W. 
Koshibae, K. Ando, S. Maekawa, and E. Saitoh, {\it Observation of the 
spin Seebeck effect}, Nature {\bf 455}, 778 (2008). 

\bibitem{WegIEEE} J.-E. Wegrowe, Q. A. Nguyen, T. Wade, {\it 
Measuring entropy due to spin-transfer}, IEEE Trans-Mag In press.

\bibitem{BauerTEP} M. Hatami, G.E.W. Bauer, Q.F. Zhang, P.J. Kelly, 
{\it Thermoelectric effects in magnetic nanostructures},
Phys. Rev. B {\bf 79}, 174426 (2009).

\bibitem{Saslow} W. M. Saslow {\it Spin pumping of current in 
non-uniform conducting magnets}, Phys. Rev. B {\bf 76}, 184434 (2007).

\bibitem{Kovalev} A. A. Kovalev and Y. Tserkovnyak, {\it 
Thermoelectric spin transfer in textured magnets},
Phys. Rev. B {\bf 
80}, 1000408(R) (2009).

 \bibitem{PRB08} J.-E. Wegrowe, S. M. Santos, M.-C. Ciornei, H.-J. 
Drouhin and M. Rubi, {\it Magnetization reversal driven by spin 
injection: a diffusive spin-transfer effect}, Phys. Rev. B, 174408 {\bf 77}  (2008).

\bibitem{Raikher} Y. L. Raikher and V. I. Stepanov, {\it Nonlinear 
dynamic susceptibilities and field-induced birefringence in magnetic 
particle assemblies}, Adv. Chem. Phys. 
{\bf 129}, 419 (2004).

\bibitem{Brown} W. F. Brown Jr., {\it thermal fluctuations of a single-domain
particle} Phys. Rev. {\bf 130}, 1677 (1963).


\bibitem{Gilbert} T. L. Gilbert, Phys. Rev. {\bf 100}, 1243 
(1955) (Abstract), reprint in {\it A phenomenological theory of 
damping in ferromagnetic materials}, IEEE Trans. Mag. {\bf 40}, 3443 
(2004).

\bibitem{Coffey} W. T. Coffey, Yu. P. Kalmykov and J. T. Waldron,
{\it The Langevin equation},
      World Scientific Series in contemporary Chemical Physics Vol. 11,
1996.


\bibitem{LevyFert} S. Zhang,  P. M. Levy, and A. Fert, {\it Mechanism 
of spin-polarized current-driven magnetization switching} Phys. Rev. 
Lett. {\bf 88}, 236601 (2002).

\bibitem{Levy1} A. Shpiro, S. Zhang, and P M. Levy, {\it Self-consistent 
treatment of nonequilibrium spin torques in magnetic multilayers}, Phys. 
Rev B {\bf 67}, 104430 (2003).

\bibitem{Levy2} J. Zhang, S. Zhang, V. Antropov, and P. M. Levy {\it 
identification of transverse spin currents in noncollinear magnetic 
structures}, Phys. Rev. Lett. {\bf 93}, 2566002 (2004).

\bibitem{Levy3} J. Zhang, P. M. Levy {\it layer bu layer approach to
transport in noncollinear magnetic structures}, Phys. Rev. B {\bf 71}, 
184426 (2005).

\bibitem{Dugaev} V. K. Dugaev, J. Barnas, {\it Classical description 
of current induced spin-transfer torque in multilayer structures}, J. 
Appl. Phys. {\bf 97} 023902 (2005).

\bibitem{BarnasFert} J. Barnas, A. Fert, M. Gmitra, I. Weymann, V. K. 
Dugaev, {\it Macroscopic description of spin transfer torque}
Mat. Science Eng. B, {\bf 126} 271 (2006).
 
\bibitem{Chung} N. L. Chung, M. B. A. Jalil, S. G. Tan, J. Guo, and 
S. Bala Kumar, {\it A study of spin relaxation on spin transfer 
switching of a noncollinear magnetic multilayer structure}, J. Appl. Phys. {\bf 104}, 084502 (2008).


\bibitem{Braatas} A. Braatas, G. E. W. Bauer, P. J. Kelly, {\it 
Non-collinear magnetoelectronics}, Phys. 
Report, {\bf 427}, 157 (2006).

\bibitem{Waintal} X. Waintal, E. B. Myers, P. W. Brouwer, D. C. Ralph, {\it Role of 
spin-dependent interface scattering in genereting current-induced 
torque in magnetic multilayers}, Phys. Rev. B {\bf 62}, 12317 (2000).

\bibitem{PRB03} J.-E. Wegrowe, {\it magnetization reversal and two 
level fluctuations by spin injection in a ferromagnetic metallic layer"},
Phys. Rev. B {\bf 68}, 214414 (2003).

\bibitem{Mazur} A statistical justification of this expression
of the chemical potential was given by P. Mazur, Physica A {\bf
261}, 451 (1998).



\bibitem{Otani} Kimura, T., Otani, Y. and Hamrle, J. ,{\it "Switching
magnetization of nanoscale ferromagnetic particle using nonlocal spin
injection"} Phys.  Rev.  Lett.  {\bf 96}, 037201 (2006).


\bibitem{Serpico} C. Serpico, I. D. Mayergoyz, G. Bertotti, M. D'Aquino, R. Bonin, 
{\it Current induced magnetization dynamics in 
nanomagnetis}  Physica
B {\bf 403}, 282


\end{thebibliography}
\end{document}